\documentclass[onecolumn,10pt,times]{elsarticle}

\usepackage{amssymb}
\usepackage{amsthm}
\usepackage{lineno,hyperref}
\usepackage{amsmath}
\usepackage{color}

\newtheorem{theorem}{Theorem}[section]

\newtheorem{definition}[theorem]{Definition}

\begin{document}

\begin{frontmatter}

\title{Two improvements of the foliation based quad meshing method}
\author[addr1]{Xiaopeng Zheng\corref{correspondingauthor}}
\cortext[correspondingauthor]{Corresponding author}
\ead{zhengxp@dlut.edu.cn}
\author[addr1]{Hao Wang}
\author[addr2]{Peng Zheng} 
\author[addr1]{Zhihui Qi}
\author[addr1]{Zhijun Cheng}
\author[addr1]{Ru Lin}
\author[addr1]{Jiayou Zhao}

\address[addr1]{Dalian University of Technology, Dalian, 116620, China}
\address[addr2]{CAEP Software Center for High Performance Numerical Simulation, Beijing, 100088, China}

\begin{abstract}
Quadrilateral meshes with high level structure and feature preserving property benefit industrial applications the most. Generation of such quad mesh remains a challenge. Quad meshes generated using surface foliation have the highest level structure, however they lack of the feature preserving ability. In this paper, we analyze the boundary curvature with Gauss-Bonnet theorem to determine whether a boundary rectangle corner preserving foliation based method exists. When it exists, we adopt a modified double cover technique together with surface foliation method to generate a corner feature preserving quad mesh. The experiments demonstrate the efficacy of our algorithm.
\end{abstract}


\end{frontmatter}


\section{Introduction}

Quadrilateral mesh generation is an essential issue in the many area. Structure level and feature preserving ability of a quad mesh matter a lot, especially in the numerical simulation area.

Surface foliation based method compute the holomorphic quadratic differentials which can induce a cylinder-decomposed high level structural quad mesh\cite{IMR1}. This method assigns zero Gauss curvature for all the boundary points of the surface, hence it yields quad meshes which have no singularities along the boundaries.

However in practice, to generate some singularities for the boundary corners whose Gauss curvature differ a lot from 0 is clearly more appropriate. 

In this research note, we proposed a frame to extend the foliation based method to corner feature preserving method. First, we analyse the Gauss curvature of the boundaries of the surface to decide that which kind singularities need to be put on the corners, and this gives the summation of the boundary curvature for the target quad mesh. Then we can compute the interior curvature summation of the target quad mesh based on Gauss-Bonnet theorem. The interior curvature summation helps to determine whether surface foliation method can be used since it can only generate valence 6 singularities. In the end, we modified the double cover technique and foliation method to achieve a corner feature preserving quad mesh generation algorithm.

Quad meshes generated by our algorithm have the highest level of structure and can also preserve the corner feature of the surface boundaries.

The structure of this research note is as follows. We review the foliation based mesh generation method and double cover technique in section \ref{preliminaries}. Section \ref{algorithms} describes our algorithms in detail and give the experiment results.

\section{Preliminaries}
\label{preliminaries}
In this section, we briefly review the foliation based quad mesh generation method. Refer to these article \cite{lei1,lei2,IMR1} for more detailed information.

\subsection{Holomorphic Quadratic Differential}

\begin{definition}[Holomorphic Differentials] Suppose $S$ is a Riemann surface. Let $\Phi$ be a complex differential form, such that on each local chart with the local complex parameter $\{z_\alpha\}$,
$\Phi = \varphi_\alpha(z_\alpha) dz_\alpha^n$, where $\varphi_\alpha(z_\alpha)$ is a holomorphic function. When $n=2$, $\Phi$ is 
called a holomorphic quadratic differential.
\end{definition}

A point $z_i \in S$ is called a \emph{zero} of $\Phi$, if $\varphi(z_i)$ vanishes. For any point away from zero, we can define a local coordinates $\zeta(p) := \int^p \sqrt{\varphi(z)}dz$, which is the so-called \emph{natural coordinates} induced by $\Phi$. The holomorphic quadratic differential $\Phi$ also defines a flat metric on the surface with cone singularities, $d_\Phi := d\zeta d\bar{\zeta}$. The curves with constant real natural coordinates are called the \emph{vertical trajectories}, with constant imaginary natural coordinates \emph{horizontal trajectories}. The trajectories through the zeros are called the \emph{critical trajectories}.

Trajectories of holomorphic quadratic differentials can from a quad mesh \cite{IMR1}.

\subsection{Surface foliation}

Surface foliation is lower dimension decomposition, decomposing the surface into one-dimensional curves. The curve is called a leaf.
If each leaf of the measured foliation $(\mathcal{F},\mu)$ is a finite loop, then $\mathcal{F}$ is called a \emph{finite measured foliation}.

Surface foliation can be computed by using the graph-valued harmonic map \cite{lei1,lei2}.

We assign an edge weight to each edge of a graph to get a metric graph.
\begin{definition}[Metric Graph] A graph $G=(V,E)$ is a one dimensional simplicial complex with a vertex set $V$ and an edge set $E$. A Riemannian metric $\mathbf{d}:E\to \mathbb{R}$ is assigned to each edge $e\in E$. $(G,\mathbf{d})$ is called a metric graph.
\end{definition}

\begin{definition}[Graph-valued Harmonic Map] Suppose $(S,\mathbf{g})$ is a surface with a Riemannian metric $\mathbf{g}$, $(G,\mathbf{d})$ is a metric graph. The mapping $\varphi: (S,\mathbf{g})\to (G,\mathbf{d})$ is harmonic, if it minimizes the harmonic energy in the homotopy class.
\end{definition}

The pre-images of the nodes of the graph are the critical trajectories, pre-images of other points on the graph induce the leaves of a foliation.

Hubbard and Masure \cite{HM79} proved the relationship between measured foliation and holomorphic quadratic differentials.
Wolf \cite{Wolf96onrealizing} showed that the holomorphic quadratic differential $\Phi$ can be obtained by the harmonic map from the Riemann surface to the metric graph.

These theorems pave the way to compute the regular quad-meshes.

\section{Algorithms}
\label{algorithms}

In this section, we focus on the genus 0 surface with multiple boundaries and give the corner feature preserving quad mesh
generation method.

\subsection{foliation based method}
without considering the corner feature, the algorithm takes genus 0 open surface as input, and output a structural quad mesh. The pipeline can be summarized as follows: 
\begin{itemize}
    \item Construct a star graph whose edge number equals to the boundary number of the open surface, and assign edge weight to obtain metric graph;
    \item Compute graph-valued harmonic map to get surface foliation;
    \item Using hodge decomposition to obtain holomorphic quadratic differential;
    \item Integrate the differential and compute the trajectories to get the quad mesh;
\end{itemize}

Foliation generated using this algorithm naturally aligns the boundaries, hence the quad mesh generated has no singularities on the boundaries.

\begin{figure}[h]
\centering
\begin{tabular}{cc}
\includegraphics[width=0.42\textwidth]{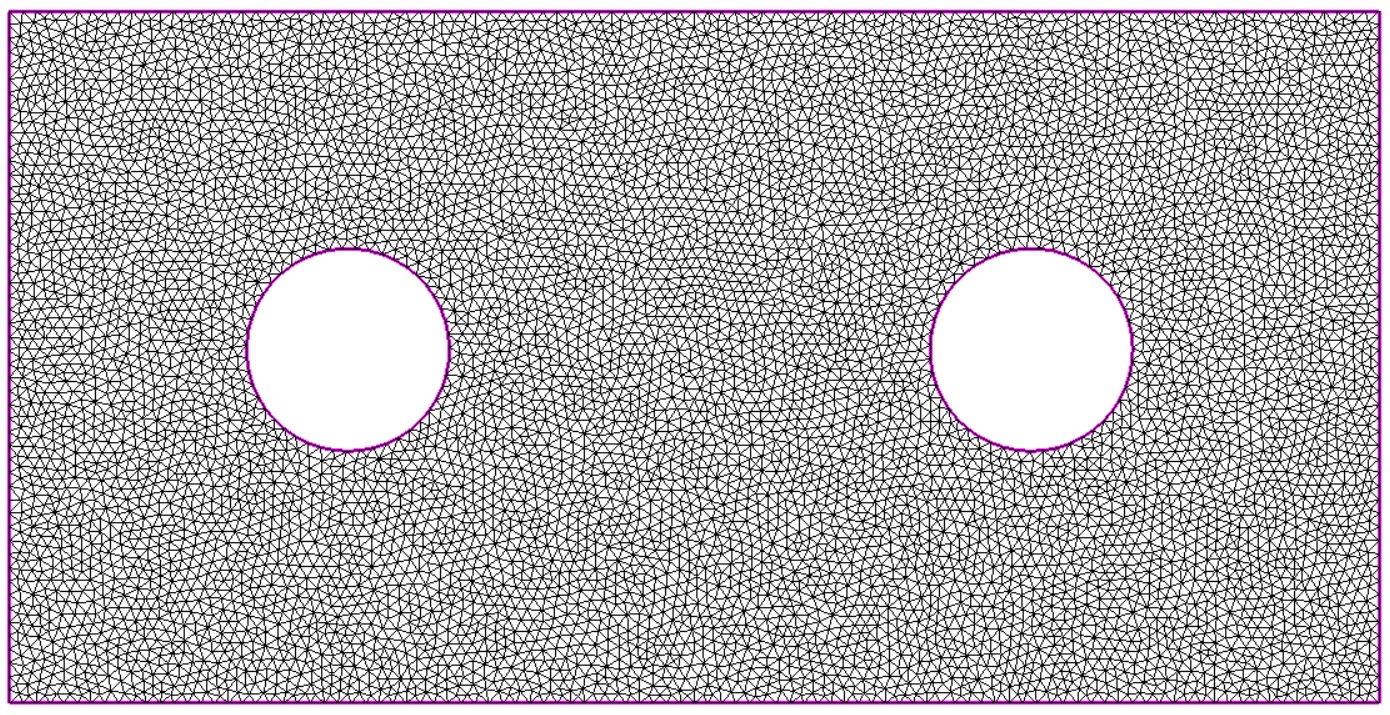}&
\includegraphics[width=0.42\textwidth]{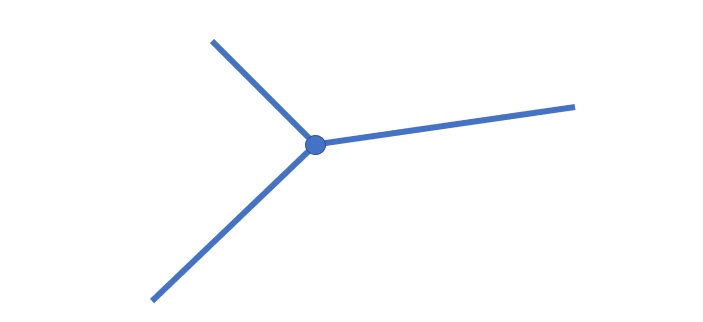}\\
(a) 3 boundaries genus 0 surface & (b) metric star graph \\
\includegraphics[width=0.42\textwidth]{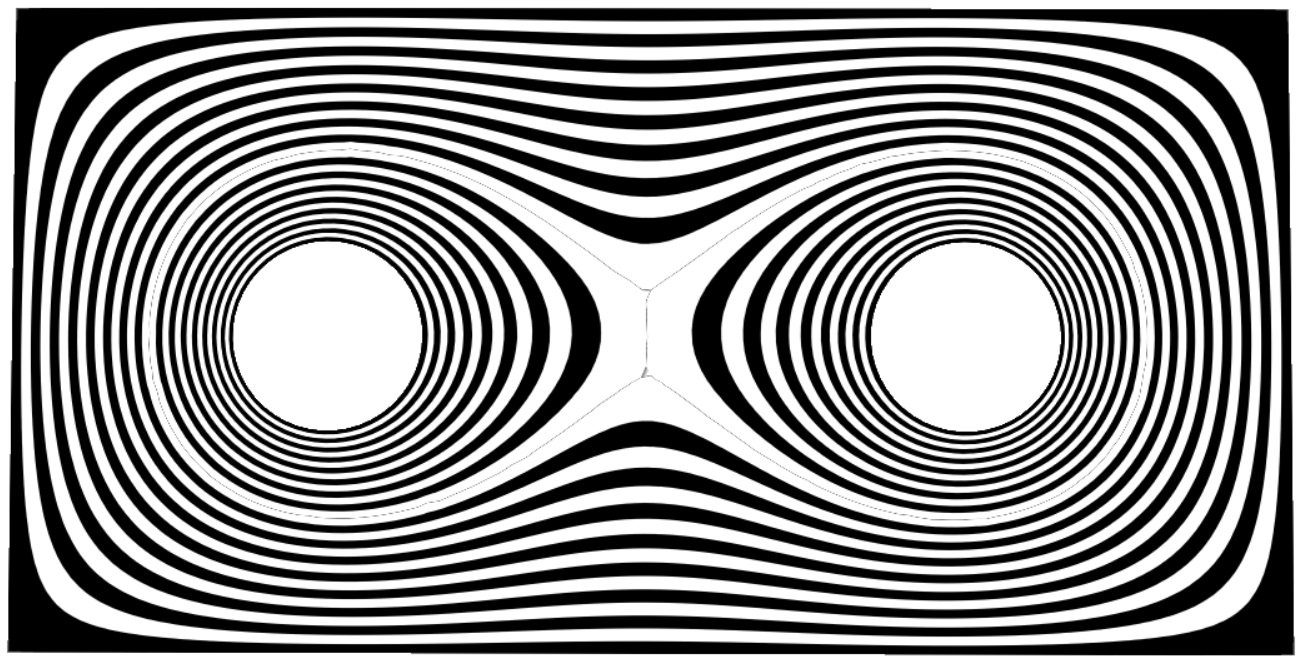}&
\includegraphics[width=0.42\textwidth]{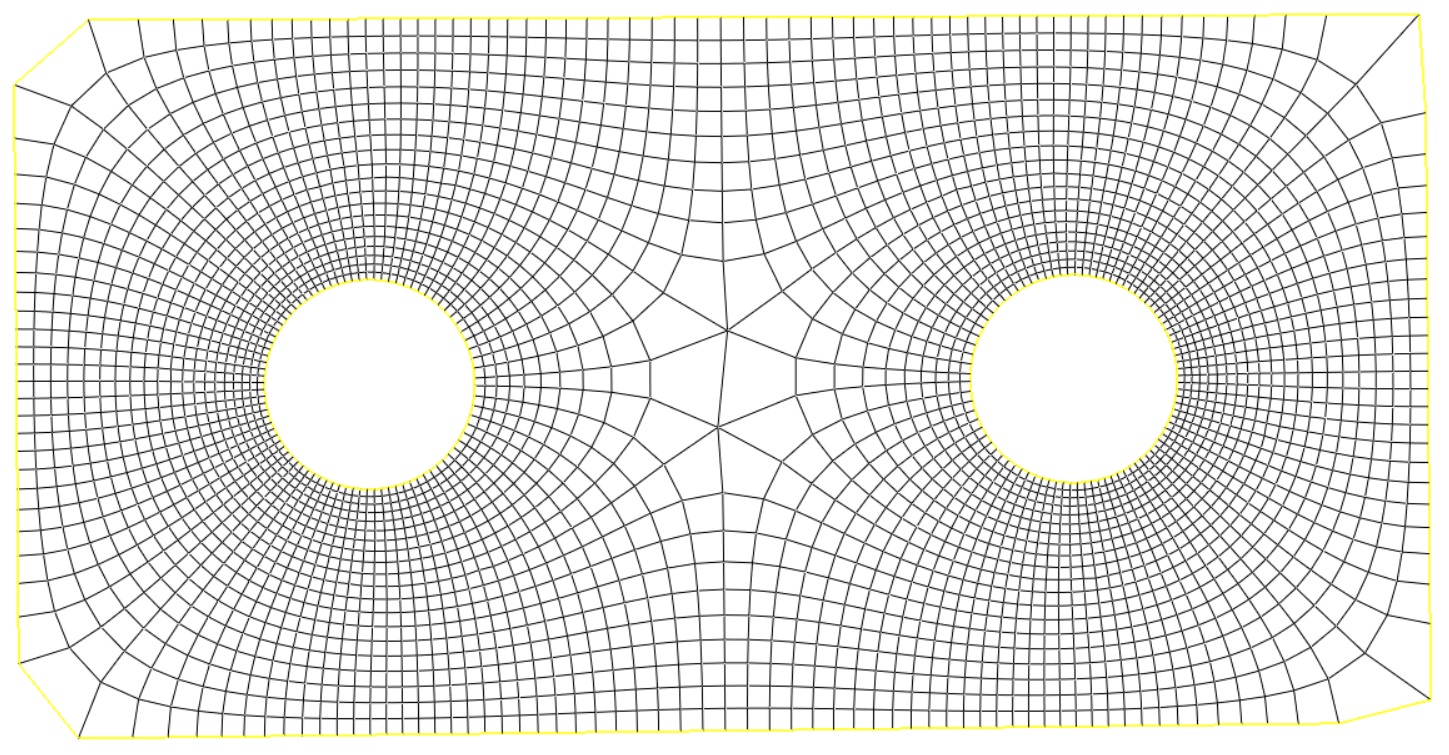}\\
(c) surface foliation & (d) quad mesh \\
\end{tabular}
\caption{quad mesh without corner feature for genus 0 surface with 3 boundaries.}
\label{fig:r1c2}
\end{figure}

For a genus 0 surface with 3 boundaries as shown in Fig.\ref{fig:r1c2}(a), we construct a star graph \ref{fig:r1c2}(b). The second row in Fig.\ref{fig:r1c2} shows the foliation and quad mesh.

\begin{figure}[h]
\centering
\begin{tabular}{cccc}
\includegraphics[width=0.22\textwidth]{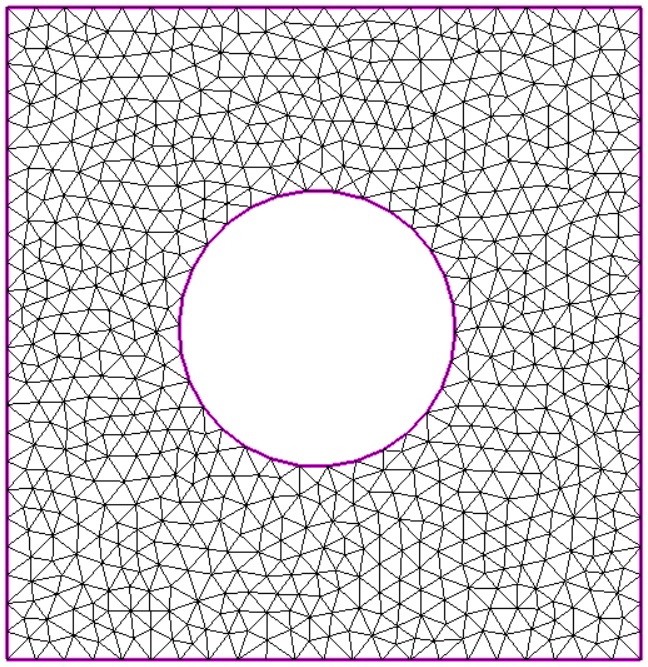}&
\includegraphics[width=0.1\textwidth]{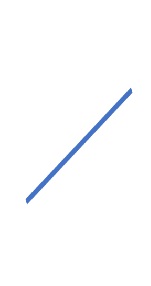}&
\includegraphics[width=0.22\textwidth]{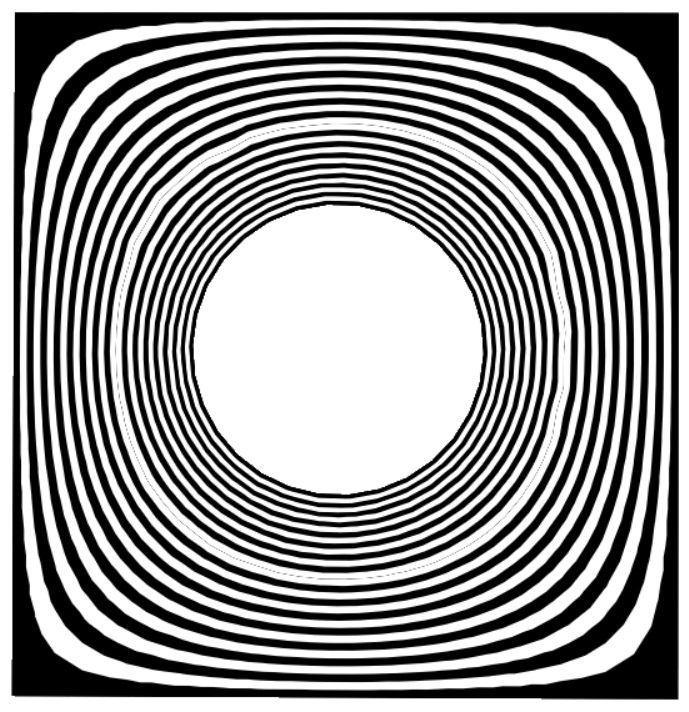}&
\includegraphics[width=0.22\textwidth]{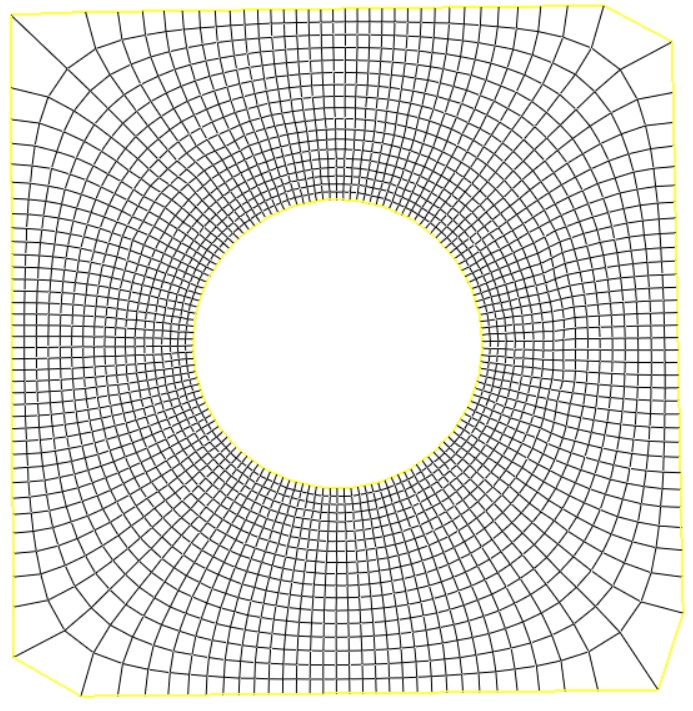}\\
\end{tabular}
\caption{quad mesh without corner feature for genus 0 surface with 2 boundaries.}
\label{fig:r1c1}
\end{figure}

Fig.\ref{fig:r1c1} shows the result of topological cylinder. 

It's easy to notice that the outer boundaries of these two examples above both have four
rectangle corners and to generate only one quad cell on each corner is more appropriate 
than two cells.

\subsection{corner preserving foliation based method}

Double cover technique can be used to convert an open surface to a close surface.
Make a copy of the input open surface and reverse the normal, then 
glue these two open surfaces along their boundaries to get a close surface.

Here we modify the double cover technique a little bit. We still make a copy of the 
input open surface and reserve its normal, then glue along only part of the boundaries
rather than all the boundaries.

\begin{figure}[h]
\centering
\includegraphics[width=0.8\textwidth]{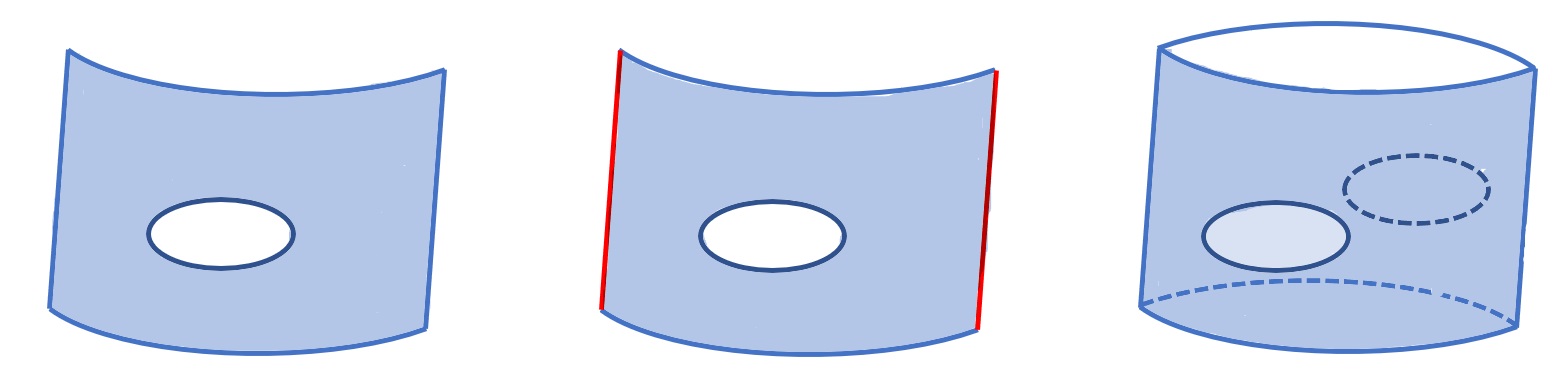}
\caption{modified double cover.}
\label{fig:cover}
\end{figure}

For a topological cylinder shown in  Fig.\ref{fig:r1c1}, double cover
technique could yield a torus. However to glue the two surfaces only along the red part of the 
boundaries returns a genus zero surface with four boundaries as show in Fig.\ref{fig:cover}.

\begin{figure*}[h]
\centering
\begin{tabular}{cc}
\includegraphics[width=0.28\textwidth]{figs/r1c1/r1c1.jpg}&
\includegraphics[width=0.3\textwidth]{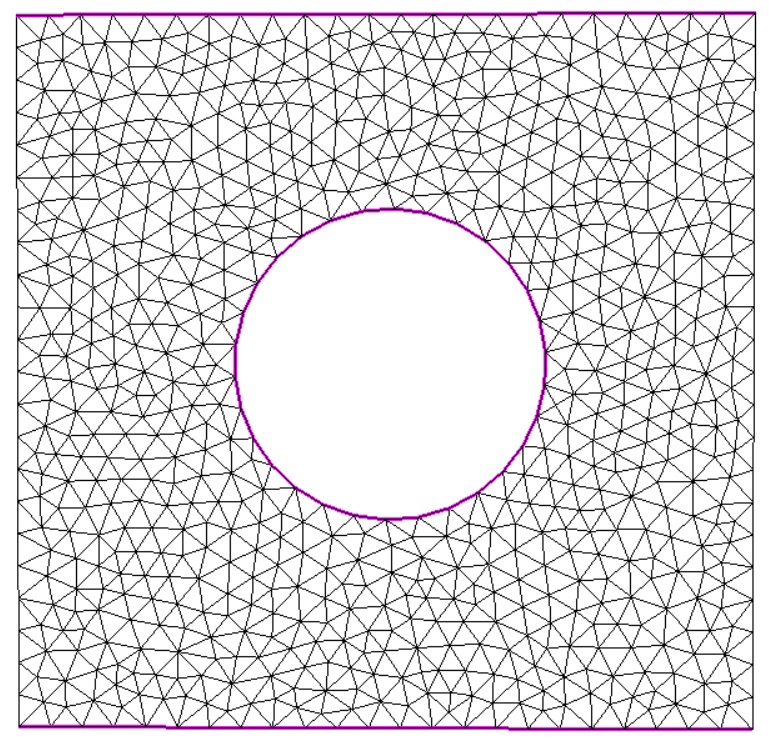}\\
(a) 2 boundaries genus 0 surface & (b) modified double cover\\
\includegraphics[width=0.3\textwidth]{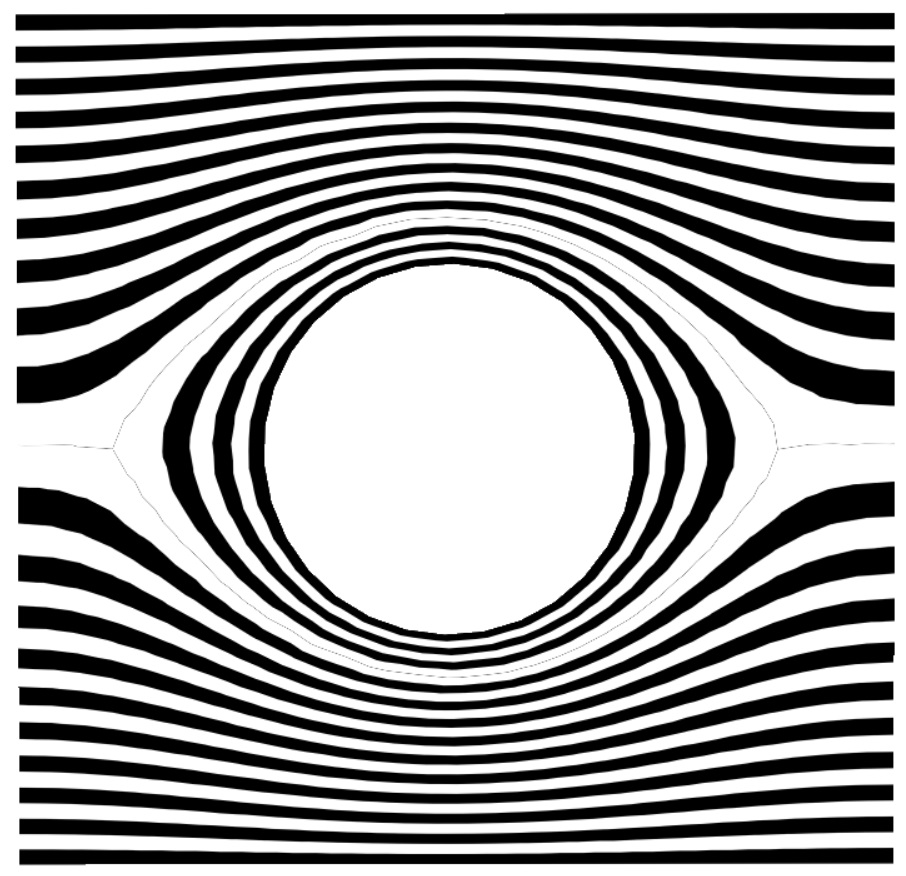}&
\includegraphics[width=0.3\textwidth]{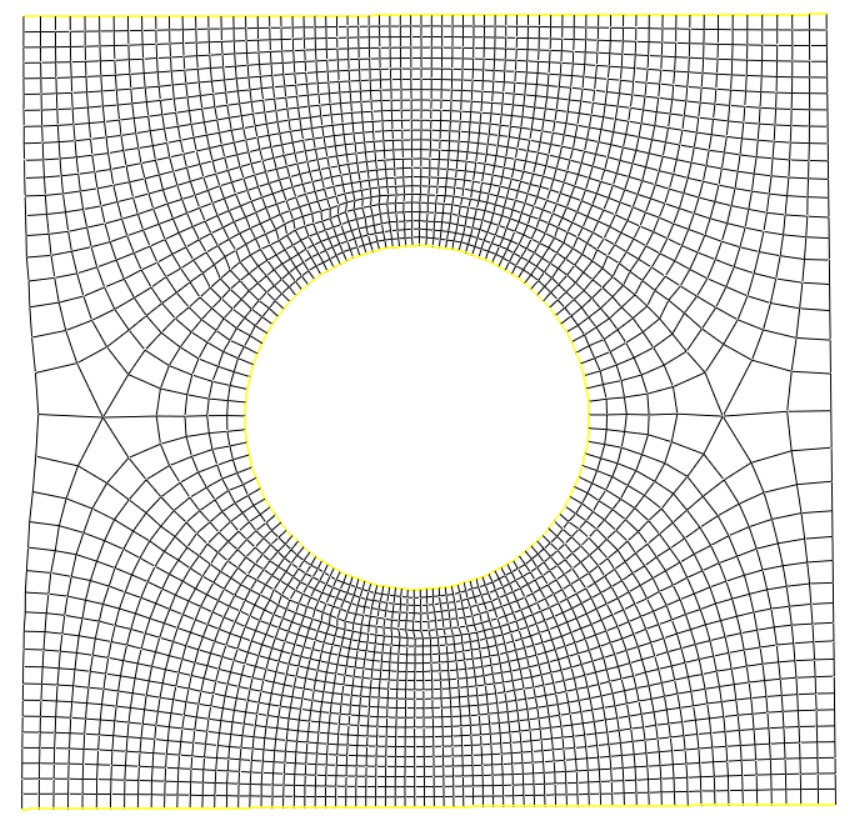}\\
(c) foliation & (d) quad mesh \\
\end{tabular}
\caption{Corner preserving method for Genus 0 surface with 2 boundaries.}
\label{fig:c_r1c1}
\end{figure*}

Now apply the surface foliation based method to this genus 0 surface with 4 boundaries and 
a different foliation and quad mesh can be generated as shown in Fig.\ref{fig:c_r1c1}. 
We can see that the quad mesh preserves the corners of the original input surface.

Hence to choose one appropriate way to glue the surface and the copy one together
may produce a corner preserving quad mesh generation method. 

\begin{figure}[h]
\centering
\includegraphics[height=0.27\textwidth]{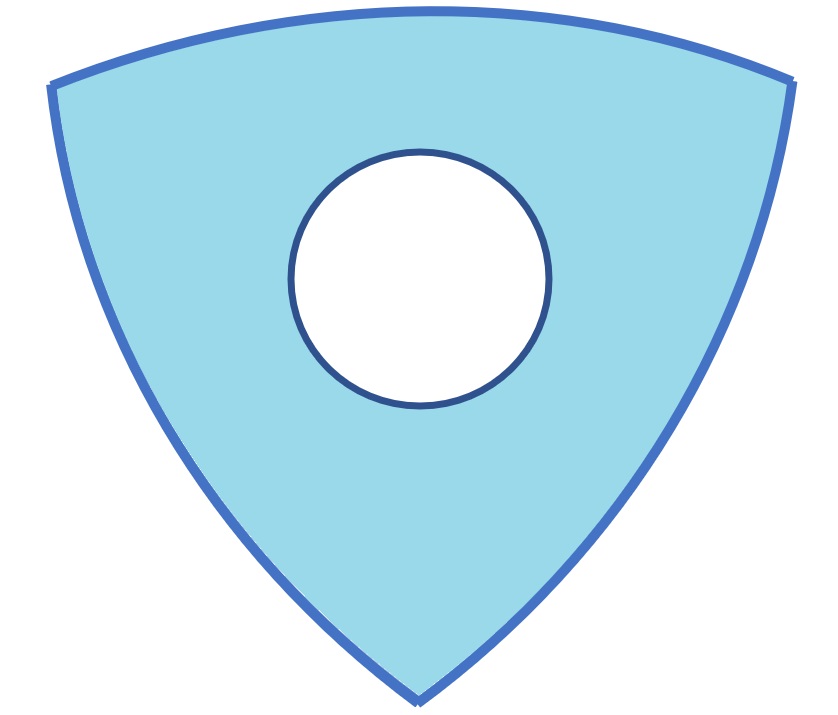}
\includegraphics[height=0.27\textwidth]{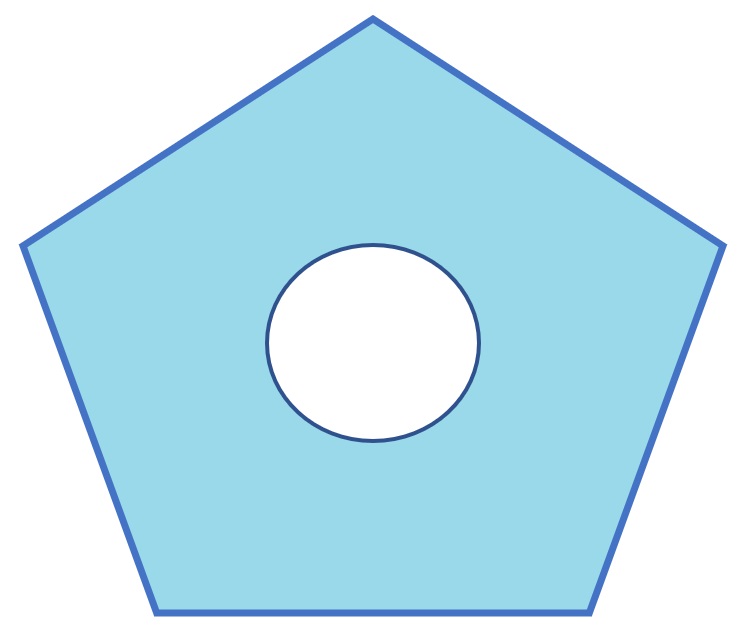}
\caption{Counterexample}
\label{fig:excep}
\end{figure}

However there exist models for which this idea fails.
Take the model in Fig.\ref{fig:excep} for example, 
we can not find one appropriate modified double cover 
and our corner preserving method does not work out.
Hence a criterion is needed.

Gauss-Bonnet theorem can be used to determine whether the corner preserving method
is valid for the given model. 
First, we analyse the boundary curvature of the input model
and assign a target Gauss curvature $\frac{n\pi}{2}$ for each 
boundary point where $n$ is a integer. Suppose summation of the target Gauss curvature
of the boundary points is $C_b$, then the summation of target Gauss curvature of the interior points 
is $2\pi\lambda-C_b$ based on Gauss-Bonnet theorem where $\lambda$ is the 
Euler characteristic. Since foliation based method can only generate valence 6 singularities whose Gauss curvature is $\pi$, only when the quotient $\frac{2\pi\lambda-C_b}{\pi}$ is an integer, can we 
use the foliation based method. It's easy to check that the quotients are both not integers for models in Fig.\ref{fig:excep}.

Based on the analysis above, the pipeline of the corner feature preserving foliation based method can 
be summarized as follows. This algorithm take a genus 0 surface with $n$ boundaries as input and output
            a corner preserved quad mesh.

\begin{itemize}
    \item Step 1: For each boundary point, compute the Gauss curvature $C_p$ and assign target 
            Gauss curvature $\frac{n\pi}{2}$. $n=1$ when $C_p>\frac{\pi}{3}$, $n=0$ when 
            $-\frac{\pi}{3}<C_p<\frac{\pi}{3}$, $n=-1$ when $-\frac{2\pi}{3}<C_p<-\frac{\pi}{3}$, 
            and $n=-2$ when $C_p<-\frac{2\pi}{3}$.
    \item Step2: compute the summation of the target Gauss curvature of the boundary points,
            denote as $C_b$. Check whether the quotient $\frac{2\pi\lambda-C_b}{\pi}$ is a integer where $\lambda$ is the 
Euler characteristic. when it's true, go to Step 3. When
            it's false, return that the foliation based method does not work for this model.
    \item Step3: divide each boundaries of the model into several segments using the corner points.
            Tag half of the segments which are all disconnected to each other and glue the surface 
            and its normal-reversed copy along these tagged segments.
    \item Step4: construct a star graph whose edge number equals to the boundary number of the covered open surface, and assign edge weight to obtain metric graph;
    \item Step5: compute graph-valued harmonic map to get surface foliation;
    \item Step6: using hodge decomposition to obtain holomorphic quadratic differential;
    \item Step7: Integrate the differential and compute the trajectories to get the quad mesh, and return half of the quad mesh as the result of the input surface.
\end{itemize}

\begin{figure}[h]
\centering
\includegraphics[width=0.4\textwidth]{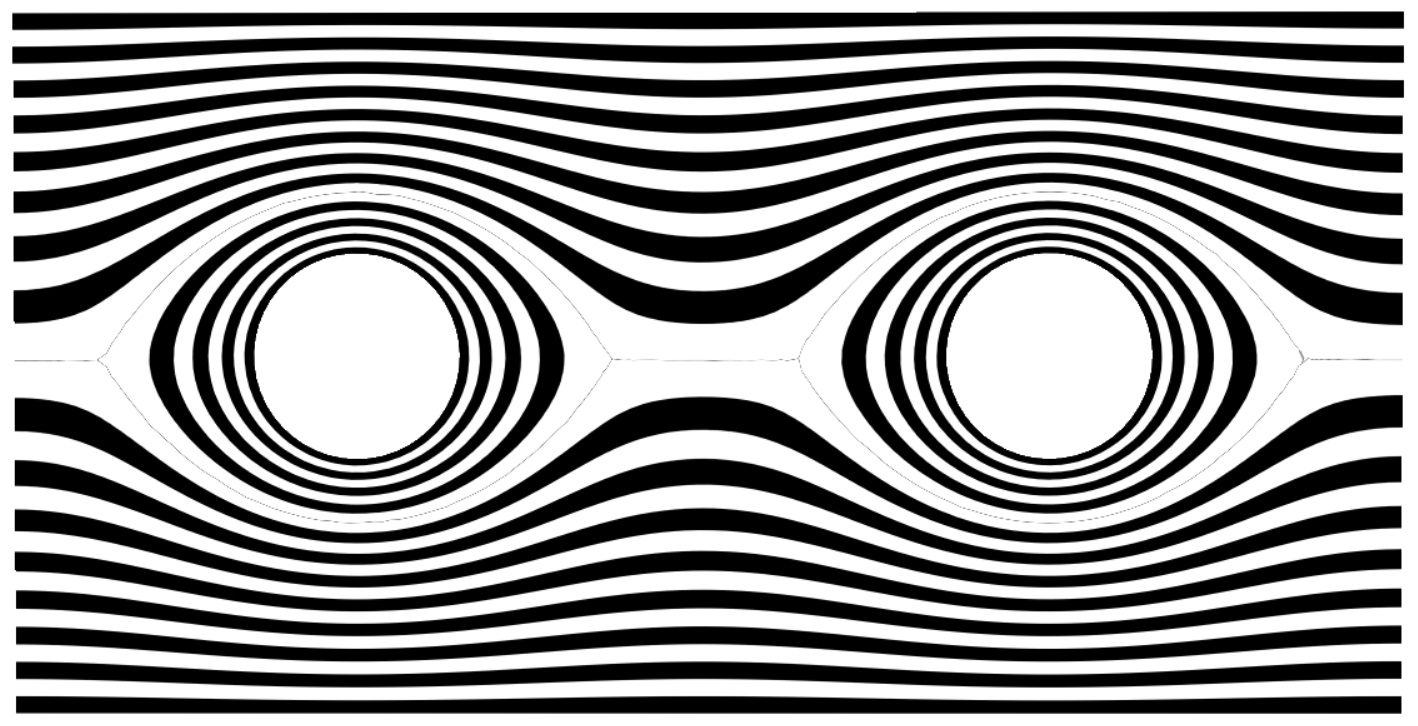}
\includegraphics[width=0.4\textwidth]{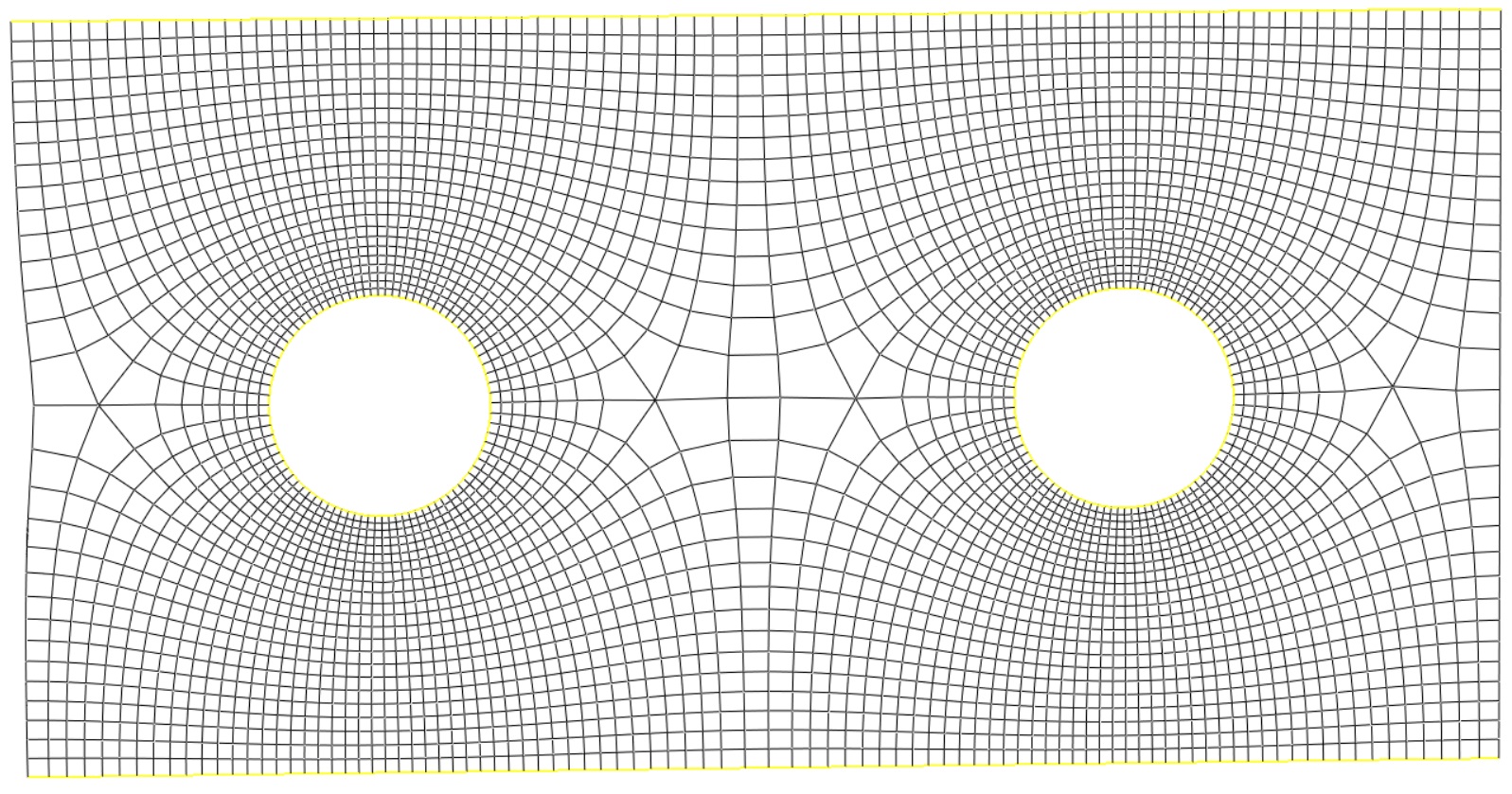}
\caption{genus 0 surface with 3 boundaries}
\label{fig:c_r1c2_quad}
\end{figure}

For the model in Fig.\ref{fig:r1c2}, the foliation and quad mesh generated by our method can preserve the corner feature as show in Fig.\ref{fig:c_r1c2_quad} and the quality is much better than the quad mesh in Fig.\ref{fig:r1c2}.

\begin{figure}[h]
\centering
\includegraphics[height=0.3\textwidth]{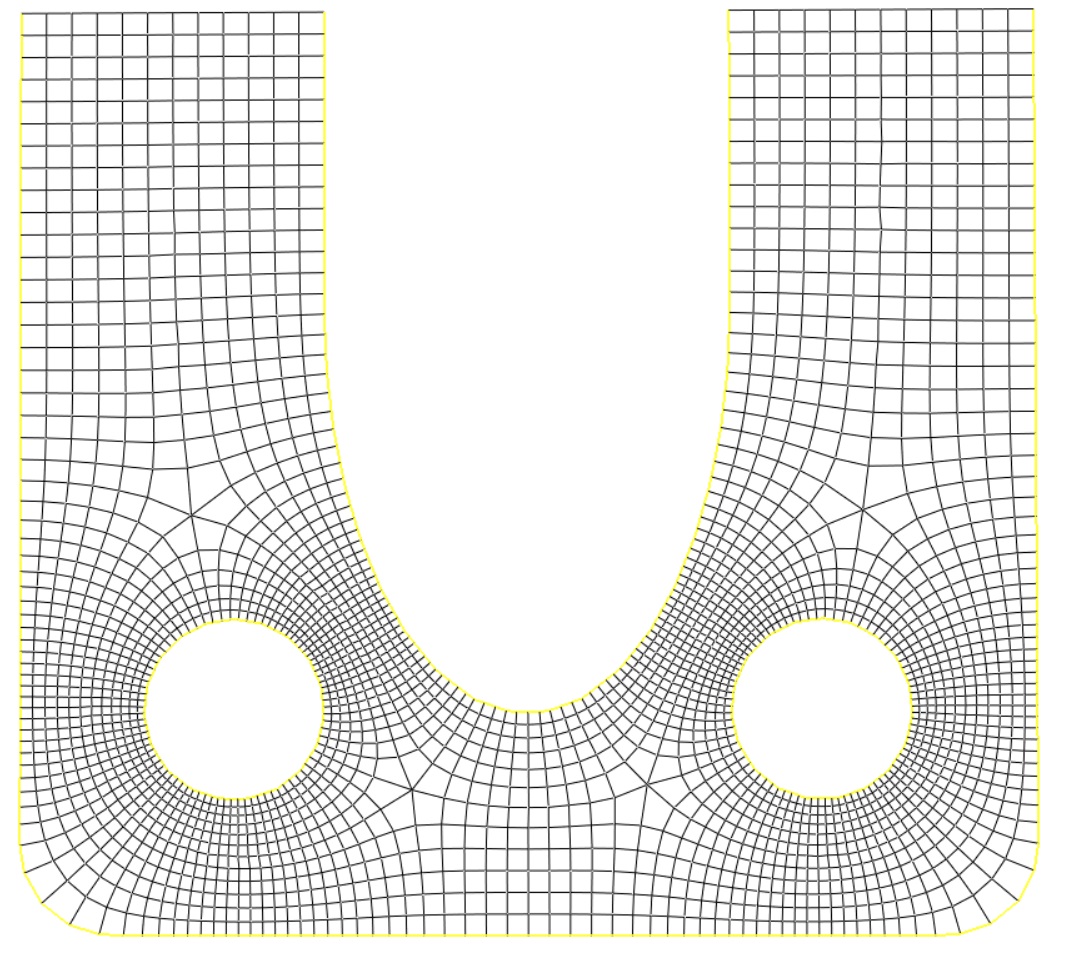}
\includegraphics[height=0.3\textwidth]{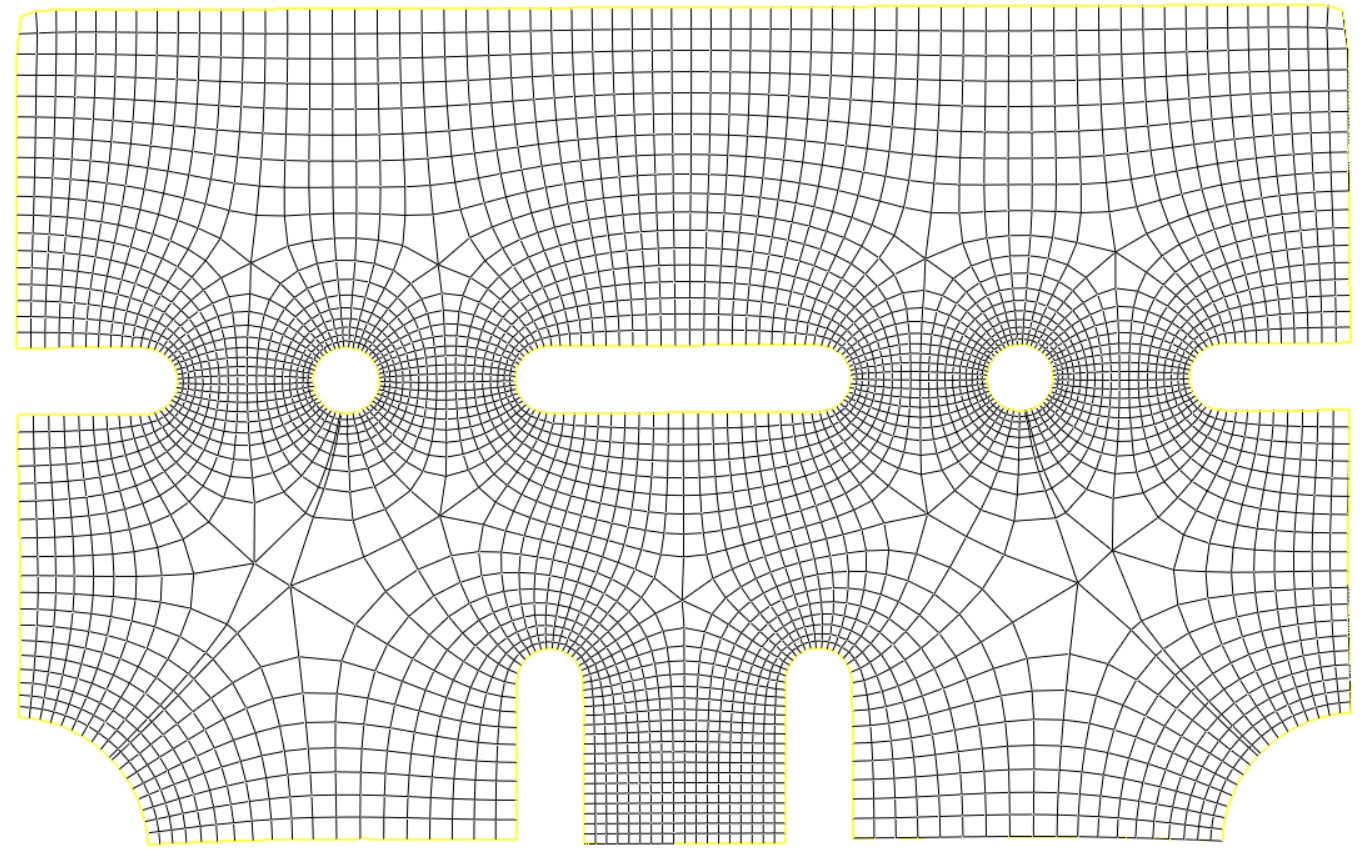}
\caption{other results}
\label{fig:c_quad}
\end{figure}

Fig.\ref{fig:c_quad} shows other results of our algorithm and they all
preserve the corner feature.

\section{Conclusion}
\label{conclusion}

In this research note, we extend the foliation based quad mesh generation method for genus zero surface 
with multiple boundaries. The quad mesh generated by our algorithm can preserve the boundary corner 
feature and have the highest structure level. The algorithm first analyse the boundary and give a 
criterion whether foliation based corner feature preserving method exists, then combine with
the modified double cover technique with foliation based method to generate a corner preserved quad mesh. 
The experiments demonstrate the efficacy of our algorithm. 
In the future, we will improve the algorithm to preserve corners for all the open surfaces.

\section*{Acknowledgement}

This work is supported by  National Natural Science Foundation of China under Grant No. 61907005, 61720106005, 61772105, 61936002.

\bibliographystyle{plain}
\bibliography{bibliography}

\end{document}